# Micromagnetic analysis of magnetic noise in ferromagnetic nanowires

**Jungbum Yoon and Chun-Yeol You**

***Department of Physics, Inha University, Incheon 402-751, Korea***

**Younghun Jo and Seung-Young Park**

***Division of Materials Science, Korea Basic Science Institute, Daejeon 305-333, Korea***

**Myung-Hwa Jung**

***Department of Physics, Sogang University, Seoul 121-742, Korea***

We investigate the magnetic thermal noise in magnetic nanowires with and without a domain wall by employing micromagnetic simulations. The magnetic thermal noise due to random thermal fluctuation fields gives important physical quantities related with the magnetic susceptibility. We find that the resonance frequency of a domain wall is distinguishable from one of a magnetic domain itself. For the single domain without a domain wall, the resonance frequency is well described by the Kittel's formula considering a ferromagnetic specimen as a simple ellipsoid with demagnetizing factors for various wire widths and thicknesses. However, additional resonance frequencies from the magnetic domain wall show the different dependences of the wire width and thickness. It implies that the spins inside the domain wall have different effective fields and the spin dynamics.

PACS number: 05.45.-a, 75.40. Gb., 75.70. kw.

Keywords: magnetic noise, thermal fluctuation, resonance frequency

Email: cyyou@inha.ac.kr, Fax: +82-32-872-7562

## I. INTRODUCTION

Since the spin-transfer torque (STT) effect [1, 2] allows the control of magnetization direction by the spin-polarized current in nanostructures, it is raising to the possibility of novel device applications with the STT effect, such as the current induced magnetization switching [3, 4], magnetic domain wall (DW) motion [5-12], magnetization precession [13, 14], and spin wave nucleation [15]. Among the applications, the race-track memory (RM) related to the current induced DW motion is attractive as a next generation nonvolatile memory device which can substitute for the current hard disk drive. The RM has advantages of the high density compared with the hard disk drive and is suitable for the mobile memory due to the nonvolatility. Furthermore, it has no mechanical part being a weakness of the conventional hard disk drive as a mobile storage [9]. The RM is operated by reading and writing of a bit data by moving the DW with the STT in ferromagnetic nanowires. There are many reports about the successful DW motion by the STT [6-10], however, the velocity and reproducibility is far from satisfactory [7, 8], and the high critical current density causes unwanted Joule heating problems [11, 12]. The underlying physics of the current induced DW motion has not yet been properly investigated. Most of the studies related with the current induced DW motion are focused on the observation of the DW motion itself, by the magnetic domain imaging [6] or the transport measurements [7, 8]. In order to elucidate the interaction between localized non-collinear spins in the DW and the conducting polarized electrons by STT, the basic information of the spins in the DW is essential and important for the efficient control of the DW.

The magnetic thermal noise due to the thermal magnetization fluctuations is related with the magnetic susceptibility by the fluctuation-dissipation theorem [16]. From the magnetic susceptibility, we can extract the important physical information such as the magnetization, anisotropy energy, and damping parameters [17]. In addition, the noise measurement is also connected with the STT effects [18-21]. However, to the best of our knowledge, there is no study about the noise spectra from the DW

[22].

In this study, we present the magnetic noise due to thermal fluctuations by employing micromagntic simulations in ferromagnetic nanowires. We find that the noise spectra of the DW have a resonance frequency distinguished from that of the single domain. We confirm that the resonance frequencies of the single domain are well agreed with the Kittel's formula for the nanowires with various widths and thicknesses. However, the resonance frequency of the DW cannot be explained by a simple model, and it implies the effective field of the spins in the DW is quite different from that of the single domain.

## II. MICROMAGNETICS

For studying on magnetic noise in ferromagnetic nanowires, we performed the micromagnetic simulation with object oriented micromagnetic framework (OOMMF) [23]. The spin dynamics is described by the Landau-Lifshitz-Gilbert (LLG) equation,

$$d\vec{M}/dt = -|\gamma|(\vec{M}\times\vec{H}_{eff}) - \alpha/M_s(\vec{M}\times d\vec{M}/dt). \quad (1)$$

Where $\gamma$ , $\alpha$ , $M_s$ , and $\vec{H}_{eff}$ are the gyromagnetic ratio, Gilbert damping constant, saturation magnetization, and the effective magnetic field, respectively. The effective magnetic field is defined as $\vec{H}_{eff} = -\mu^{-1}\partial E/\partial\vec{M}$ . The energy density $E$ is a function of $\vec{M}$ specified by Brown's equations [24], including magnetic anisotropy, exchange, magnetostatic, and Zeeman energies due to the applied magnetic field. Here, the archetypal LLG equation doesn't consider the motion of spin due to the random thermal fluctuation field, which always exists in the finite temperature and it is the source of the magnetic thermal noise. In order to include the thermal noise effect, we employed the stochastic LLG equation with the thermal fluctuation term,

$$d\vec{M}/dt = -|\gamma|\vec{M}\times(\vec{H}_{eff} + \vec{h}_{fl}(t)) - \alpha/M_s(\vec{M}\times d\vec{M}/dt). \quad (2)$$

Here $\vec{h}_{fl}(t)$ is a highly irregular field term due to the Gaussian stochastic process. $\vec{h}_{fl}(t)$ has the statistical properties, $\langle h_{fl,i}(t)\rangle = 0$ and $\langle h_{fl,i}(t)h_{fl,j}(s)\rangle = 2D\delta_{ij}\delta(t-s)$ , where $i$ and $j$ are Cartesian indices, the constant $D$ measures the strength of the thermal fluctuations. Dirac $\delta$ represents that above finite temperature the autocorrelation time of $\vec{h}_{fl}(t)$ is much shorter than the rotational-response time of the system, while the Kronecker $\delta$ represents that the different components of $\vec{h}_{fl}(t)$ are assumed to be uncorrelated [25].

Micromagnetic simulations are calculated for a magnetic structure divided by small cells. If a cell size is small enough, the simulation is accurate but it takes longer simulation time. So the cell size must be optimized by considering the accuracy and simulation time. Recently, it has been reported that the proper renormalization process is required for the physical quantities such as the saturation magnetization [26]. Due to the finite cell size which is much larger than the atomic size, the contribution of smaller wavelength magnon than the cell size cannot be properly considered in micromagnetic simulations. To avoid this problem in the finite temperature, the renormalization process is necessary. In this work, however, we perform the simulation without the renormalization process at 300 K. We pay our attention to a qualitative behavior of the magnetic noise spectra and the resonance frequency, not a quantitative analysis. The quantitative results are not changed without the renormalization process in our study.

For micromagnetic simulations, we select a model system of a 10-nm thick, 80-nm wide, and 1000-nm long NiFe (permalloy, Py) nanowire. A tail-to-tail transverse DW is placed at center position with a small notch ($5\times10$ nm$^2$), as shown in Fig. 1 (a). The notch traps the domain wall as a potential well, but the effect on the resonance frequency is negligible when the size of notch is not large. The material parameter of Py is as follows: the saturation magnetization $M_s = 8.6\times10^5$ A/m, the exchange stiffness $A_{ex} = 13\times10^{-12}$ J/m, and the Gilbert damping constant $\alpha = 0.01$. We perform our simulations with a zero applied magnetic field and take a cell size of $5\times5\times5$ nm$^3$. Also, the thickness and width of the Py nanowire is varied from 10 to 30 nm with 5 nm steps for a fixed 80 nm width and from 50 to 120 nm with 10 nm steps for a fixed 10 nm thickness, respectively. The DW structure of narrow magnetic nanowires can be one of the transverse, vortex, and asymmetric transverse types depending on the thickness, width, and saturation magnetization due to the minimization of the demagnetization energy [27]. We limited the DW type as a symmetry transverse type by choosing the proper ranges of the thickness and width in our study. Because of a spatial symmetry of the vortex type DW, the sum of magnetic noise from the whole vortex DW structure cancels each other and gives very small noise signal, while the transverse type DW has better results. We obtained the spin configurations in the Py nanowires at each temporal moment for sufficiently long time ($10^{-7}$ sec). The random thermal fluctuation field term is updated with the interval of $10^{-14}$ sec. The magnetization configurations are stored $10^{-11}$ sec steps. The resonance frequency due to the magnetic noise can be extracted by the fast Fourier transform (FFT) from the time varying the transverse magnetization component ($M_y$).

## III. RESULTS AND DISCUSSION

The resonance frequency due to magnetic fluctuations in the Py nanowires is calculated by employing the OOMMF with the stochastic LLG equation [28]. First, we consider the 80-nm wide, 10-nm thick, and 1000-nm long Py nanowire with a transverse type DW configuration as shown in Fig. 1 (a). The temporal variation of the total $M_y$ and the results of the FFT power of the total $M_y$ is shown in Fig. 1 (b) and (c), respectively. The random thermal motion of the magnetization due to the finite temperature is clearly shown in Fig. 1 (b). We find two resonance frequencies at 7.1 and 9.4 GHz, as shown in Fig. 1 (c). It must be noted that there is only one peak (9.7 GHz) obtained for the single domain case, as shown in Fig. 1 (d). The resonance frequency is well agreed with the Kittel's equation's result, 9.3 GHz. This equation employs the corresponding demagnetization factors of $N_x = 0.00248$, $N_y = 0.10932$, and $N_z = 0.8882$, with the relation of

$$f = \frac{\gamma}{2\pi} \times \sqrt{(H_{eff} + (N_z - N_x) \times M_s) \times (H_{eff} + (N_y - N_x) \times M_s)} \, . \qquad (3)$$

Here $N_x$, $N_y$, and $N_z$ are demagnetization factors for an corresponding ellipsoid [29]. To clarify the source of the additional peak (7.1 GHz), we employed the concept of the local spectra [22]. The local spectra are calculated from the limited region around the DW, as shown the black rectangle in Fig. 1 (a). In order to obtain the local spectra, we consider the sums of only the local $M_y$ instead of the sums of the $M_y$ for the whole region. With the local spectra analysis, we can confirm that the additional peaks are generated due to the DW. The additional peaks from the DW have distinguished resonance frequencies from that of the single domain. It implies the effective field acting on spins inside of the DW is different from the inside of the single domain. Furthermore, the resonance frequency from the DW cannot be described with Kittel's formula, because the demagnetization factor cannot be determined for the DW.

In order to reveal the behavior of the resonance frequency of the DW, we perform the same micromagnetic simulations with various widths and thicknesses. In Fig. 2, we plotted the local spectra from (a) the transverse-type DW and (b) the single domain part for various wire widths (50 to 120 nm with 10 nm steps). The decrement of the resonance frequencies with the wire-width variation is clearly observed for the domain and DW, respectively. At Fig. 3 (a), the width dependence of the resonance frequencies for the domain and DW is plotted with the calculated resonance frequencies with Eq. (3). The agreements between the results of micromagnetic simulations and Eq. (3) are excellent. The corresponding demagnetization factors are shown in Fig. 3 (b). For all width, $N_x$ ~ 0, and $N_x + N_y + N_z = 1$ are hold, Eq. (3) can be rewritten as follows:

$$f = \frac{\gamma}{2\pi} \times \sqrt{(H_{eff}^2 + M_s H_{eff} + N_y(1 - N_y) M_s{}^2)} \, . \tag{4}$$

In this equation, it is clear that the resonance frequency only depends on $N_y(1 - N_y)$ term, and the corresponding $N_y(1 - N_y)$ is depicted in the Fig. 3 (b). The relations between the resonance frequency and demagnetization factors are clear. However, it must be pointed out that even though the resonance frequency of the DW shows the same trend, decreased with broadening the wire width, the dependence of the resonance frequency of the DW is more complicated, which we will discuss later.

Next, we perform the same micromagnetic simulations with various thicknesses (from 10 to 30 nm with 5 nm steps) with a fixed width of 80 nm as shown in Fig. 4 (a) and (b). In this case, the thickness dependence of the resonance frequencies of the single domain and DW shows different behaviors. While the resonance frequencies of the single domain increase with increasing the wire thickness, those of the DW decrease. The agreements between micromagnetic simulations and Eq. (3) are still excellent as shown Fig. 5 (a). However, the resonance frequencies of the DW show the opposite behavior. In Fig. 5 (b), corresponding demagnetization factors with $N_y(1-N_y)$ are also depicted. Only the resonance frequencies of the single domain can be explained with Eq. (4) and $N_y(1-N_y)$ term, but that of the DW shows the different behavior.

Since the resonance frequency of the ferromagnetic body is a function of the effective field and the magnetization, the distinguished resonance frequency implies that the effective field inside of the DW is quite different to that of the single domain. Furthermore, the dependence of the width and thickness is different. Inside of the single domain, where the magnetizations are collinear, $\nabla \cdot \overrightarrow{M} = 0$ is satisfied. However, this condition is no longer valid inside of the DW, where the magnetization directions are abruptly changed. Furthermore, there are many free poles at the both side of the nanowire edges, it also gives additional stray field. Therefore, different effective fields for the spins inside of the DW are understandable. However, there is no simple analytic model for the realistic transverse type DW so that we cannot estimate the effective field inside of the DW. In order to obtain better understanding of the resonance frequency of the DW, experimental radio-frequency measurements must be carried out.

## IV. CONCLUSION

The magnetic noise spectra due to thermal fluctuations in the Py nanowires are investigated by the OOMMF with the stochastic LLG equation. The Kittel's formula well elucidates the resonance frequencies of the magnetic noise spectra with the variation of the wire width and thickness for the single domain. We find that the resonance frequencies of the DW are distinguished from that of the single domain and confirm by the local spectra concept. The distinguishable resonance frequency of the DW implies that the spins inside of the DW have difference effective fields from that of the single domain. We believe that the magnetic noise measurement and the analysis of the DW will give a better understanding of the spin dynamics for the spins inside of the DW.

**ACKNOWLEDGEMENT**

This research was supported by Nano R&D program (2008-02553, 2009-0082441) through the National Research Foundation of Korea(NRF) funded by the Ministry of Education, Science and Technology and the IT R&D program of MKE/KEIT (2009-F-004-01).

**REREFENCES**

## Figure Captions

Fig. 1. (a) Magnetization configuration of the Py nanowire with transverse type DW at the center. It is generated by the OOMMF simulation. The in-plane magnetization is represented by small arrows and the blue and red direction arrows. The black rectangle represents a local region for magnetic thermal noise spectra of the domain wall. (b) Random thermal motion of the magnetization due to the thermal fluctuations at 300 K. Magnetic noise spectra due to thermal fluctuations of the Py nanowire (c) with and (d) without the DW without the external magnetic field.

Fig. 2. Local magnetic noise spectra due to thermal fluctuations in (a) the transverse type DW and (b) single domain. The widths are varied from 50 to 120 nm with 10 nm steps.

Fig. 3. (a) Comparison of the resonance frequency from the magnetic noise spectra of the single domain, DW, and from Eq. (3) as a function of the width of the Py nanowires. (b) Corresponding demagnetization factors and $N_y(1-N_y)$.

Fig. 4. Local magnetic noise spectra due to thermal fluctuations in (a) the transverse type DW and (b) single domain. The thicknesses are varied from 10 to 30 nm with 5 nm steps.

Fig. 5. (a) Comparison of the resonance frequency from the magnetic noise spectra of the single domain, DW, and from Eq. (3) as a function of the thickness of the Py nanowires. (b) Corresponding demagnetization factors and $N_y(1-N_y)$.

Fig. 1.

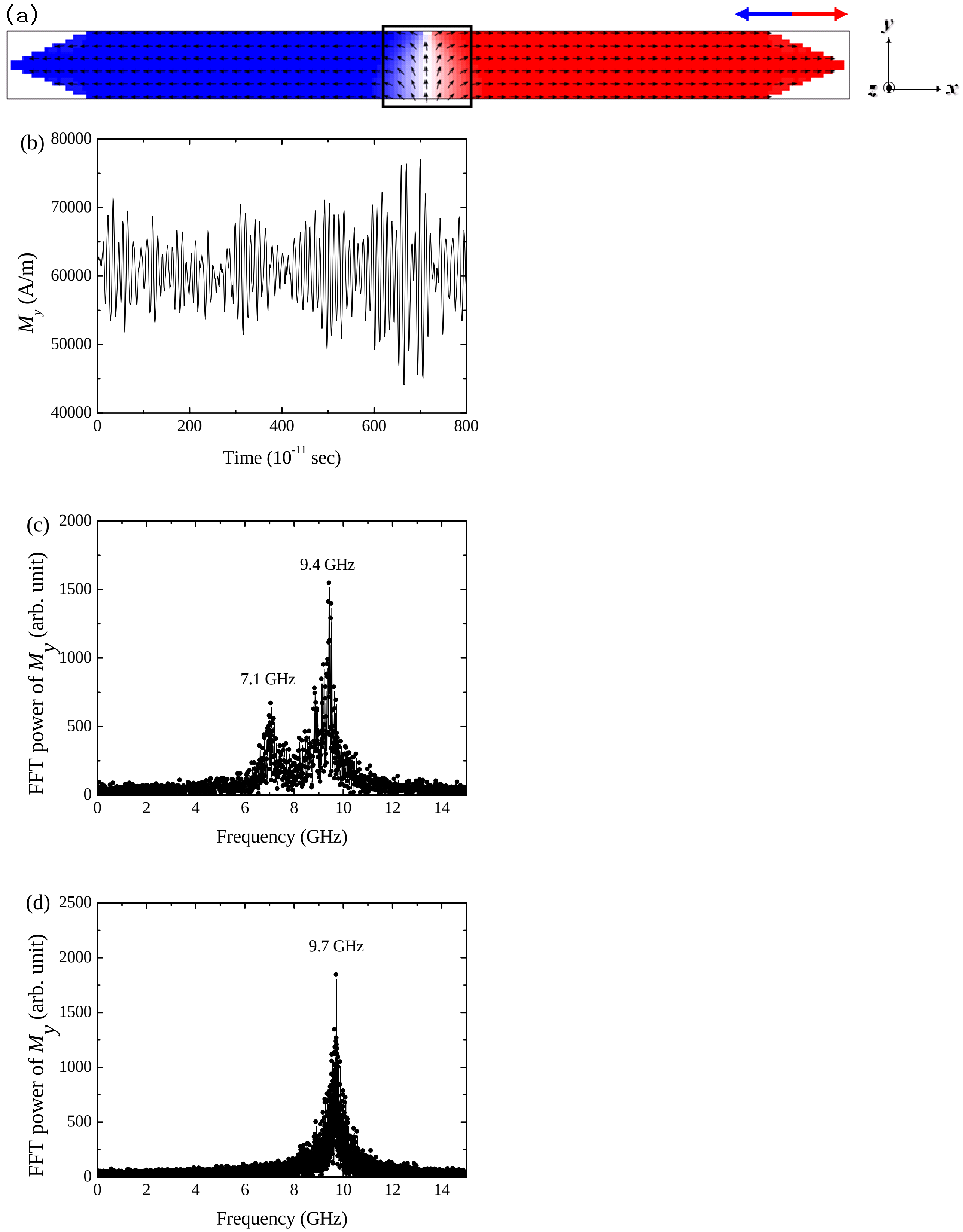

Fig. 2.

(a)

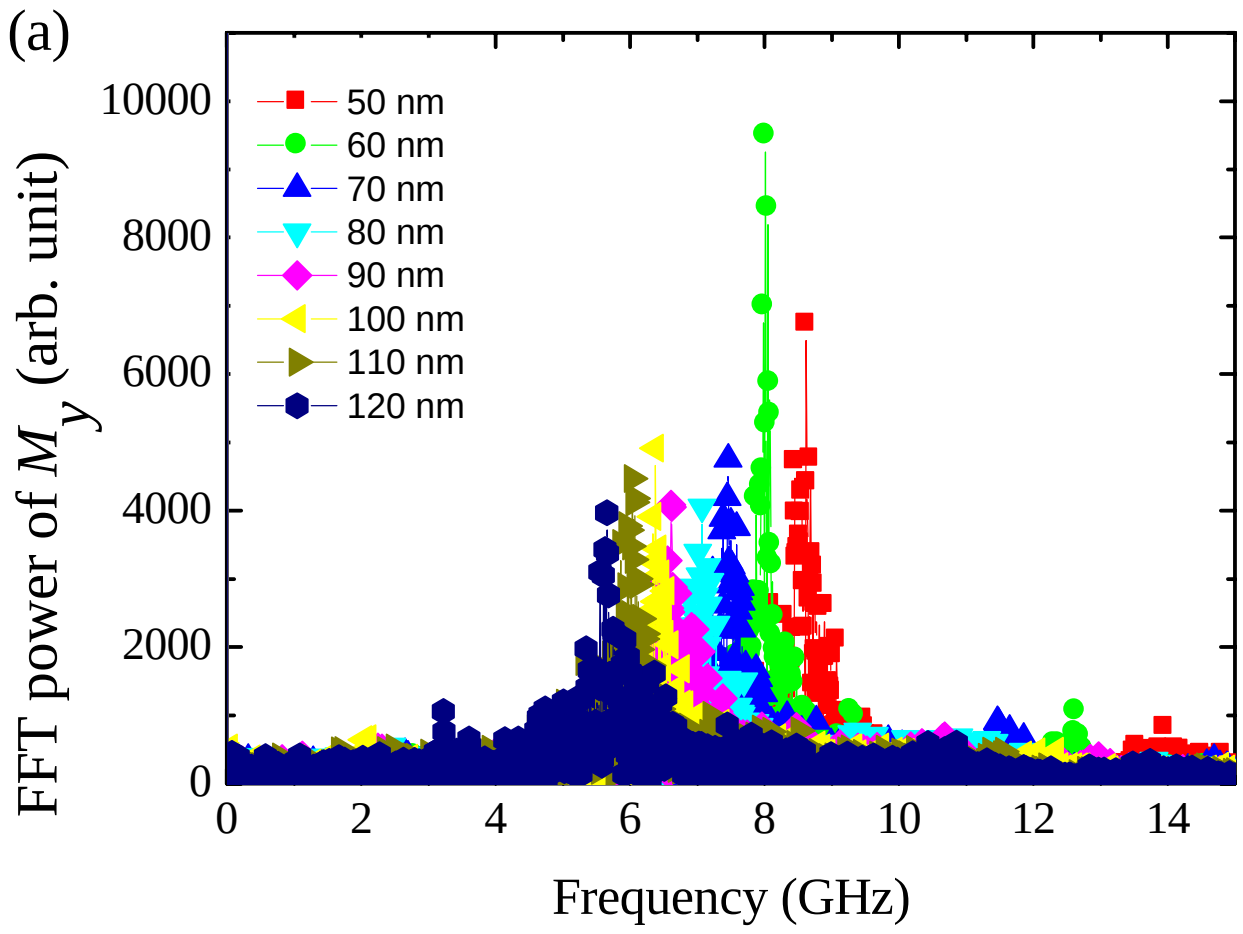

(b)

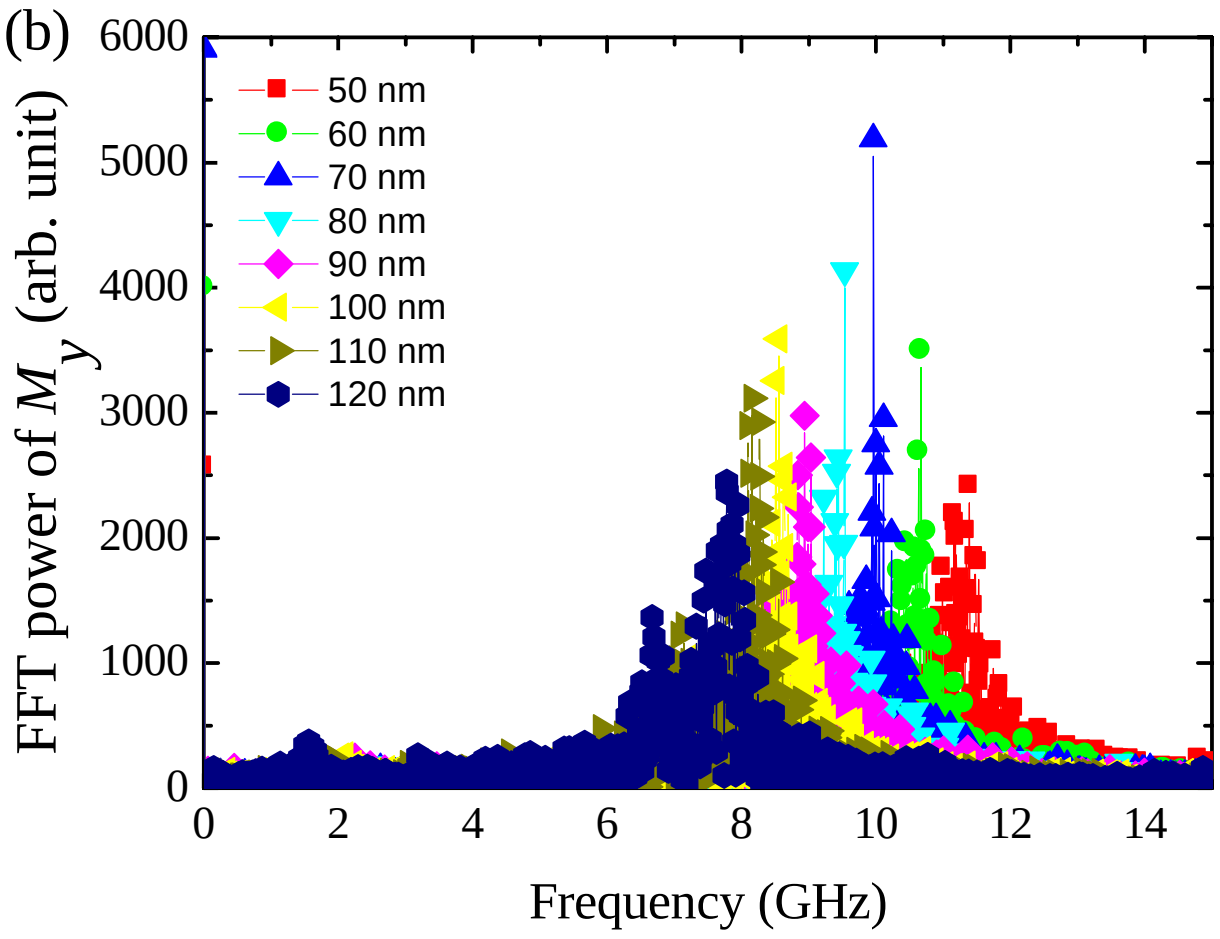

Fig. 3.

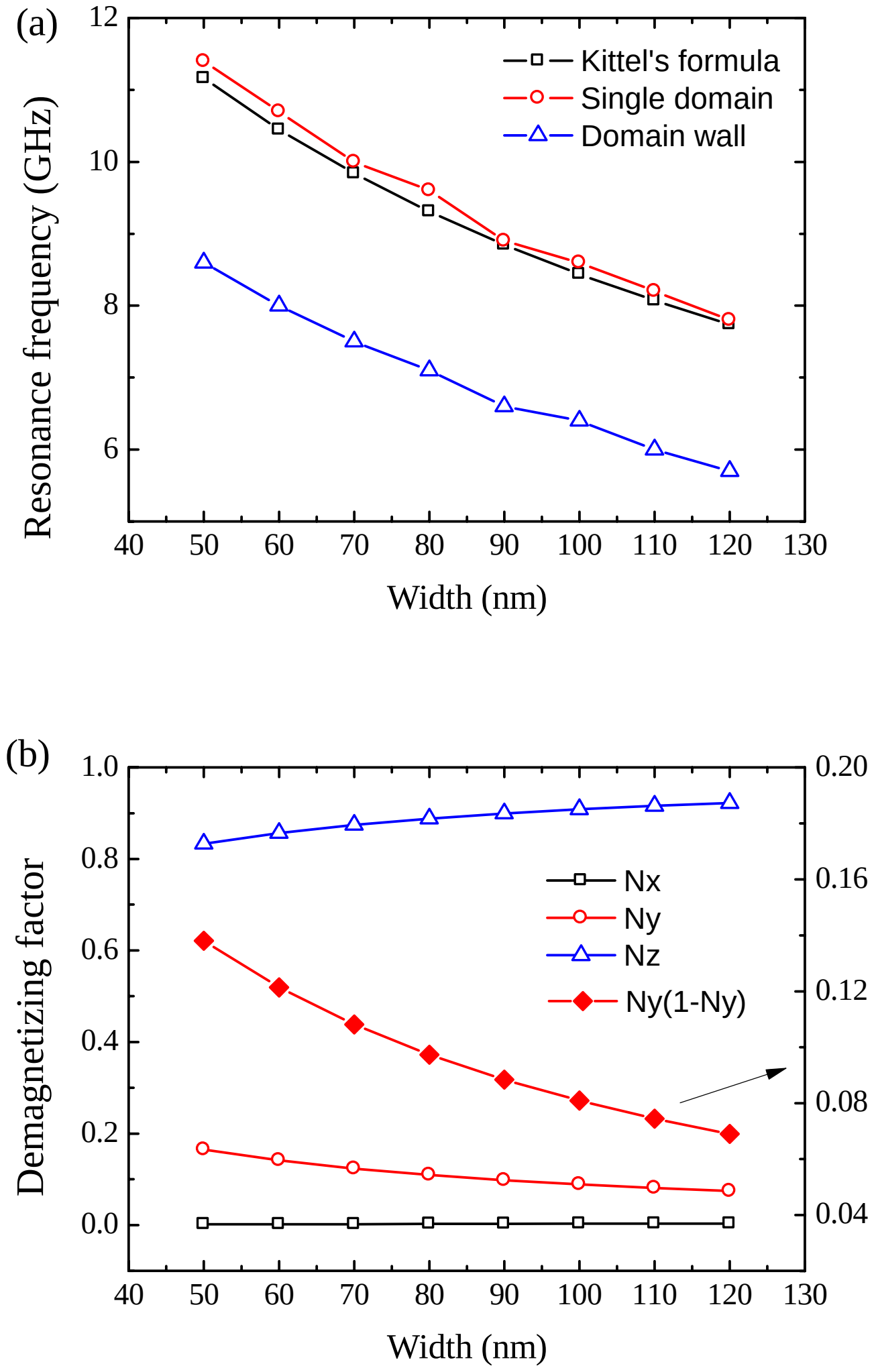

Fig. 4.

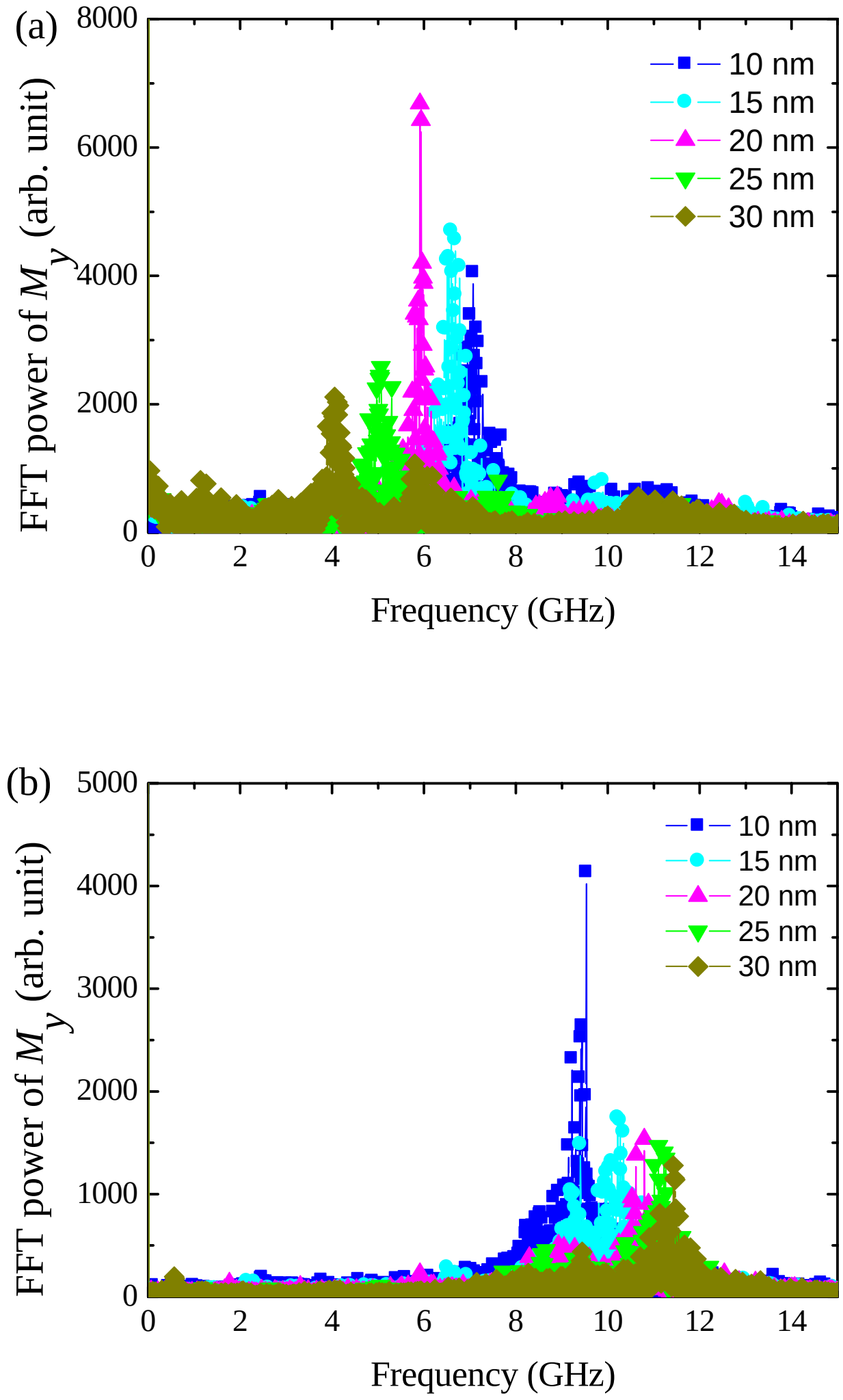

Fig. 5.

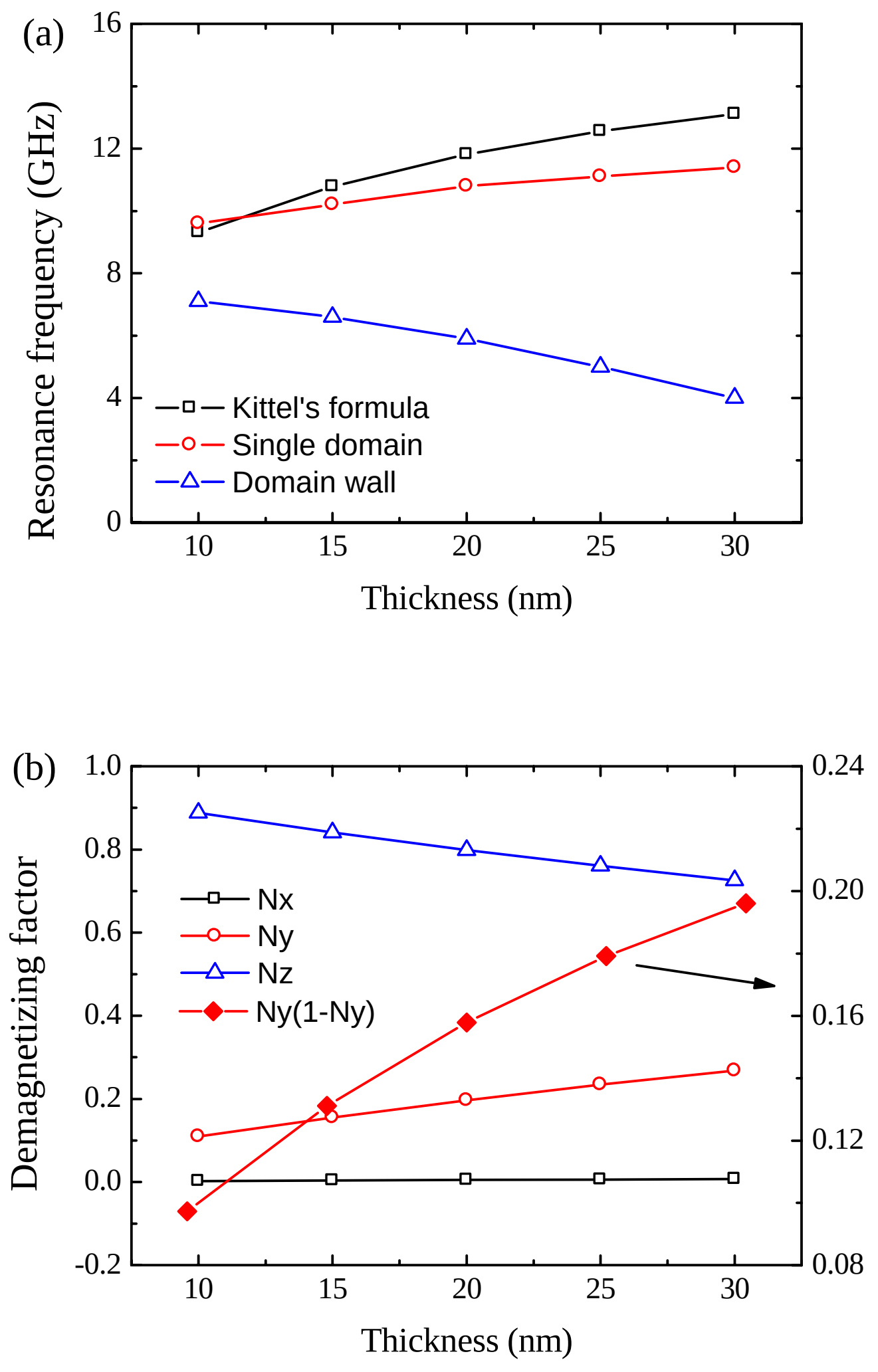